\documentclass[12pt]{article}

\usepackage{mathptmx} 
\usepackage{helvet} 
\usepackage{courier} 
\usepackage{float}
\usepackage[top=1in, bottom=1in, left=1in, right=1in]{geometry}

\usepackage{amsmath, amssymb}
\usepackage{mathtools}  
\usepackage{amsfonts}

\usepackage{graphicx}
\usepackage{color}
\usepackage{caption}
\usepackage{subcaption}

\usepackage{cite}

\usepackage{hyperref}

\title{Searching for gravitational-wave bursts from cosmic string cusps with the Parkes Pulsar Timing Array third data release}
\author{
    {\small  Yong Xia}$^{1,2,3}$, 
    \href{https://orcid.org/0000-0001-9782-1603}{\small  Jingbo Wang}$^{3,*}$,
    {\small  Sachiko Kuroyanagi}$^{4,5}$,
    \href{https://orcid.org/0000-0002-7662-3875}{\small  Wenming Yan}$^{1,6,*}$,
    {\small  Yirong Wen}$^{1,2,3}$,\\
    \href{https://orcid.org/0009-0001-5071-0962}{\small  Agastya Kapur}$^{7}$,
    \href{https://orcid.org/0000-0002-2035-4688}{\small  Daniel J. Reardon}$^{8,9}$,
    \href{https://orcid.org/0000-0002-9583-2947}{\small  Andrew Zic}$^{9,10}$,
    {\small  Jing Zou}$^{1,2,3}$,
    \href{https://orcid.org/0000-0002-0475-7479}{\small  Yi Feng}$^{11,12}$,\\
    \href{https://orcid.org/0000-0003-3432-0494}{\small  Valentina Di Marco}$^{10,13,14}$,
    \href{https://orcid.org/0009-0001-5633-3512}{\small   Saurav Mishra}$^{15}$,
    \href{https://orcid.org/0000-0002-1942-7296}{\small  Christopher J. Russell}$^{16}$,\\
    \href{https://orcid.org/0000-0003-4498-6070}{\small  Shuangqiang Wang}$^{1}$,
    \href{https://orcid.org/0000-0007-8062-1454}{\small  De Zhao}$^{1}$,
    \href{https://orcid.org/0000-0001-7049-6468}{\small  Xingjiang Zhu}$^{17,18}$\\
}

\date{}

\begin{document}

\maketitle

\noindent
\small $^{1}$ Xinjiang Astronomical Observatories, Chinese Academy of Sciences, Urumqi 830011, China\\
\small $^{2}$ University of Chinese Academy of Sciences, Beijing 100049, China\\
\small $^{3}$ Institute of Optoelectronic Technology, Lishui University, Lishui 323000, China\\
\small $^{4}$ Department of Physics, Nagoya University, Nagoya 464-8602, Japan\\
\small $^{5}$ Instituto de Fisica Teorica UAM-CSIC, Universidad Autonoma de Madrid, 28049 Madrid, Spain\\
\small $^{6}$ Xinjiang Key Laboratory of Radio Astrophysics, Urumqi 830011, China\\
\small $^{7}$ School of Mathematical and Physical Sciences, Macquarie University, NSW 2109, Australia\\
\small $^{8}$ Centre for Astrophysics and Supercomputing, Swinburne University of Technology, P.O. Box 218, Hawthorn, VIC 3122, Australia\\
\small $^{9}$ OzGrav: The Australian Research Council Centre of Excellence for Gravitational Wave Discovery, Hawthorn, VIC 3122, Australia\\
\small $^{10}$ Australia Telescope National Facility, CSIRO, Space and Astronomy, PO Box 76, Epping, NSW 1710, Australia\\
\small $^{11}$ Research Center for Astronomical Computing, Zhejiang Laboratory, Hangzhou 311100, China\\
\small $^{12}$ Institute for Astronomy, School of Physics, Zhejiang University, Hangzhou 310027, China\\
\small $^{13}$ School of Physics and Astronomy, Monash University, Clayton VIC 3800, Australia\\
\small $^{14}$ OzGrav: The ARC Center of Excellence for Gravitational Wave Discovery, Clayton VIC 3800, Australia\\
\small $^{15}$ Centre for Astrophysics and Supercomputing, Swinburne University of Technology, P.O. Box 218, Hawthorn, VIC 3122, Australia\\
\small $^{16}$ CSIRO Scientific Computing, Australian Technology Park, Locked Bag 9013, Alexandria, NSW 1435, Australia\\
\small $^{17}$ Department of Physics, Faculty of Arts and Sciences, Beijing Normal University, Zhuhai 519087, China\\
\small $^{18}$ Institute for Frontier in Astronomy and Astrophysics, Beijing Normal University, Beijing 102206, China\\
\small $^{*}$ Correspondence: 1983wangjingbo@163.com; yanwm@xao.ac.cn\\



\renewcommand{\abstractname}{\normalfont\Large\textbf{Abstract}}

\begin{abstract}
Pulsar timing arrays (PTAs) are designed to detect nanohertz-frequency gravitational waves (GWs). Since GWs are anticipated from cosmic strings, PTAs offer a viable approach to testing their existence. We present the results of the first Bayesian search for gravitational-wave bursts from cosmic string cusps (GWCS) using the third PPTA data release for 30 millisecond pulsars. In this data collection, we find no evidence for GWCS signals. We compare a model with a GWCS signal to one with only noise, including a common spatially uncorrelated red noise (CURN), and find that our data is more consistent with the noise-only model.
We then establish upper limits on the strain amplitude of GWCS at the pulsar term, based on the analysis of 30 millisecond pulsars, after finding no compelling evidence. We find the addition of a CURN with different spectral indices into the noise model has a negligible impact on the upper limits. And the upper limit range of the amplitude of the pulsar-term GWCS is concentrated between $10^{-12}$ and $10^{-11}$. Finally, we set upper limits on the amplitude of GWCS events, parametrized by width and event epoch, for a single pulsar PSR J1857$+$0943. {Moreover, we derive upper limit on the cosmic string tension as a function of burst width and compare it with previous results.} 
\end{abstract}


\section{Introduction}          
\label{sect:intro}

Low-frequency GWs between $10^{-9}$ and $10^{-7}$ Hz can be detected with high sensitivity using a pulsar timing array (PTA) which consists of a group of millisecond pulsars with extremely stable rotational periods \cite{foster1990}. Owing to their stability, it is anticipated that timing residuals affected by a gravitational wave (GW) can be observed by closely monitoring the times of arrival (TOAs) of radio pulses from these millisecond pulsars \cite{hellings1983upper}. Timing residuals are the discrepancies between the TOAs and the predictions of the pulsar timing model \cite{edwards2006tempo2}. Other than GWs can lead to timing residuals, including clock faults, unmodeled variability in the solar wind, interstellar medium scattering, flaws in the solar system ephemeris (SSE), and offsets caused by the observational instrument \cite{tiburzi2016study}. Therefore, there will be a significant difficulty in detecting gravitational waves. The North American Nanohertz Observatory for Gravitational Waves (NANOGrav) \cite{jenet2009north}, the European Pulsar Timing Array (EPTA) \cite{kramer2013european}, the Parkes Pulsar Timing Array (PPTA) \cite{manchester2013international}, and the Indian Pulsar Timing Array (InPTA) \cite{nobleson2022low} are among the PTA collaborations that are currently in operation. These partnerships result in the International Pulsar Timing Array (IPTA) \cite{verbiest2016international}. Furthermore, the first analyses from the Chinese Pulsar Timing Array (CPTA) \cite{xu2023searching} and the MeerTime Pulsar Timing Array (MPTA) \cite{miles2023meerkat} have been published recently.

PTAs are currently the most promising method for detecting nanohertz GWs, primarily generated by the inspiraling of supermassive black hole binaries (SMBHBs) \cite{burke2019astrophysics}. The incoherent superposition of these binary systems produces a stochastic GW background. Recent studies by NANOGrav \cite{agazie2023nanograv}, PPTA \cite{reardon2023search}, EPTA \cite{antoniadis2023second}, and CPTA \cite{xu2023searching} have reported that the significance of a GW origin is between 2$\sigma$ and 4$\sigma$. The observed correlations between pulsar timing residuals closely match the predicted Hellings-Downs pattern, a characteristic signature of such a background. This finding has significant implications for our understanding of supermassive black holes and the large-scale structure of the Universe.

%
Beyond the stochastic gravitational-wave background, PTA data can also be used to probe a variety of other astrophysical phenomena. For instance, continuous gravitational waves emitted by single supermassive black holes \cite{zhu2014all}, known as continuous waves, can be detected or constrained using PTA observations. 
Additionally, PTAs are sensitive to gravitational wave memory \cite{wang2015searching,agazie2024nanograv}, a permanent distortion of spacetime caused by extreme gravitational events like the merger of two black holes. Furthermore, PTAs can provide insights into the nature of dark matter \cite{porayko2018parkes}. By searching for characteristic signals from ultralight scalar-field dark matter, PTAs can constrain the properties of this elusive component of the Universe.

One particularly intriguing possibility is the detection of GWCS, hypothetical defects in the fabric of spacetime that may have formed during the early Universe. Such bursts, if they exist, could be within the detectable frequency range of PTAs.
Theoretical models suggest that a spontaneous symmetry-breaking phase transition in the early Universe \cite{damour2000gravitational} could have resulted in the formation of macroscopic, one-dimensional, stable, and highly energetic cosmic strings. Quantum field theory and condensed matter models predict these cosmic strings as topological defects, which play a role in various supersymmetric unified field theories, including D-brane models \cite{yonemaru2021searching}. Initially, cosmic strings were proposed as a potential explanation for the formation of large-scale structure in the early Universe, with symmetry breaking occurring at the grand unification scale \cite{dvali2004formation}.

Due to the predicted emission of GWCS, experiments like the PTAs offer a potential method to verify their existence. 
The most powerful GW bursts are generated at cusps, which are highly Lorentz-boosted singularities on string loops \cite{damour2000gravitational}. A cusp is a region in the loop that is highly Lorentz-boosted and produces a powerful beam of GWs. In general, the gravitational wave background is primarily the result of the overlap of multiple bursts. However, in specific regions of parameter space, a limited number of distinct GW bursts can contribute significantly to the overall background. A sufficiently large GW amplitude from such an event could render it detectable as a single burst. The Laser Interferometer Gravitational-Wave Observatory (LIGO) \cite{abbott2009first,abbott2016search,abbott2018constraints} has conducted searches for individual GW from cosmic strings in the direction of fast radio burst sources. Leveraging the second data release from the PPTA project, Yonemaru et al. \cite{yonemaru2021searching} performed the first search for GWCS, employing a frequentist-based approach.
In this paper, we will search and place limits on GWCS based on the third PPTA data release (PPTA-DR3) using a Bayesian approach. 

This paper is organized as follows. In Section \ref{sect:data}, we provide a short description of PPTA-DR3. In Section \ref{sect:signal}, we examine the impact of GWCS on TOAs from a PTA. We provide a concise overview of the Bayesian method and software employed in this search in Section \ref{sect:methods}. In Section \ref{sect:result}, we demonstrate the implementation of the proposed algorithm on the PPTA data and discuss the resulting outcomes. Finally, the conclusion is presented in Section \ref{sect:conclusions}.

\section{Data}
\label{sect:data}
This analysis is based on the PPTA-DR3 data and the noise analyses of individual pulsars. We will provide a concise summary of the most significant aspects of this data set, with additional details available in Zic et al. \cite{zic2023parkes} and Reardon et al. \cite{reardon2023gravitational}.

The 64-m Parkes "Murriyang" radio telescope was utilized to observe 32 pulsars, which are included in the release. Observations commenced in 2004, and the data were collected over the course of 18 years (MJD 53040-59640) consisting of approximately 48 hours of observations made every 2-3 weeks. This release combines an updated version of the second PPTA data release (PPTA-DR2) with approximately three years of more recent data. The data were predominantly obtained using an ultra-wide-bandwidth receiver system that operates between 704 and 4032 MHz. One of the pulsars that was previously included in PPTA-DR2, PSR J1732$-$5049, was excluded due to the significant uncertainties in its TOAs, which resulted in periodic observations ceasing in 2011. Six additional binary pulsars (PSRs J0125$-$2327, J0614$-$3329, J0900$-$3144, J1741$+$1351, J1902$-$5105, and J1933$-$6211) and one solitary pulsar (PSR J0030$+$0451) are included in this release, in contrast to previous data releases. After the UWL receiver was commissioned, these pulsars were incorporated into PPTA observations, despite not having been previously included. In large part, the enhanced observation efficiency of the UWL receiver enabled the inclusion of these pulsars in the PPTA observations.

The data set utilized for this analysis, which encompasses the pulsar ephemerides and TOAs, is detailed in Zic et al. \cite{zic2023parkes}. Commencing with the timing analyses of the preceding data releases \cite{reardon2016timing,reardon2021parkes}, the initial timing models of each pulsar were fitted using {\sc TEMPO2} \cite{edwards2006tempo2}. For the pulsars newly included in the PPTA following the release of PPTA-DR2, we adopt the initial timing models presented by Curyło et al. \cite{curylo2023wide}.

Despite the inclusion of observations from 32 pulsars in PPTA-DR3, we will analyze only 30 pulsars in this study. We excluded the globular cluster MSP PSR J1824$-$2452A because its steep-spectrum red noise is too intense for a common process to affect it. Although globular-cluster dynamics may also be involved, the disturbance is likely intrinsic to the pulsar \cite{shannon2010assessing}. This analysis, as well as the subsequent GWCS search, excludes PSR J1741$+$1351 because the data set contains only 16 unique observations of this pulsar, which is insufficient for noise process simulation.

\section{Signal of the GWCS}
\label{sect:signal}
We will examine the effect of GWCS on TOA in pulsar timing data in this section and provide an extensive overview of the signal and noise models incorporated. A GW passing through a pulsar will induce a shift in the observed rotational frequency, which may result in either an increase or decrease in the frequency. When a GW passes through the Earth, the rotational frequencies observed for all pulsars in the PTA will be affected, leading to either an increase or decrease. The timing residuals induced by a GWCS will display distinct shapes depending on the epoch and duration of the burst, yet these shapes adhere to a deterministic pattern.

The timing residuals will be introduced as a result of the discrepancy between the observed rotational frequency and the pulsar's timing model-fitted rotational frequency, regardless of the situation. Detweiler \cite{detweiler1979pulsar} provided the timing residuals induced by GWs:
\begin{equation}
    r(t)=\sum_{a=+,\times} F^a(\hat{\Omega},\hat{p}) \int^{t}\Delta h_a(\hat{\Omega},t') dt',
    \label{eq:rt}
\end{equation}
where $\hat{p}$ and $\hat{\Omega}$ are the directions of the pulsar and the GW propagation, respectively. $F^a(\hat{\Omega},\hat{p})$ is the antenna pattern defined by Anholm et al. \cite{anholm2009optimal}:

\begin{equation}
    F^a(\hat{\Omega},\hat{p})=\frac{1}{2}  \frac{\hat{p}^i \hat{p}^j}{1+\hat{\Omega}\cdot \hat{p}}e_{ij}^a(\hat{\Omega}).
    \label{eq:F}
\end{equation}
The GW polarization tensors, denoted as $e_{ij}^a(a=+,\times)$, are defined:
\begin{alignat}{2}
    e_{ij}^+(\hat{\Omega}) &= \hat{m_i}\hat{m_j}-\hat{n_i}\hat{n_j}\\
    e_{ij}^{\times}(\hat{\Omega}) &= \hat{m_i}\hat{n_j}+\hat{n_i}\hat{m_j},
\end{alignat}
where $\hat{m}$ and $\hat{n}$ are the polarization basis vectors. For ease of computation we define:
\begin{alignat}{2}
    \hat{m}&=sin\delta \hat{x}-cos\delta \hat{y}\\
    \hat{n}&=-cos\alpha cos\delta \hat{x}-cos\alpha sin\delta \hat{y}+sin\alpha \hat{z},
\end{alignat}
$\hat{x},\hat{y},\hat{z}$ are unit vectors in the Cartesian coordinate system, $(\alpha,\delta)$ is the gravitational wave source position, we can get:
\begin{equation}
    \hat{\Omega}=-sin\alpha cos\delta \hat{x}-sin\alpha sin\delta \hat{y}-cos\alpha \hat{z},
\end{equation}
$\Delta h_a(\hat{\Omega},t')$ represents the metric perturbation difference between the Earth and the pulsar, as determined by the linearly polarized GWCS. The time-domain waveform of a plus-polarized GWCS event is detailed by Yonemaru et al. \cite{yonemaru2021searching}:
\begin{align}
     h_+(t) &= 
     \begin{cases}
     A_{\text{fit}}\left[ \left\lvert t-t_0 \right\rvert^{1/3}-(\frac{1}{2}W)^{1/3} \right] &   (t_0-\frac{1}{2}W\le t <t_0+\frac{1}{2}W) \\
     0 &  (\text{otherwise}) \\
     \end{cases}
     \label{eq:hplus}\\
     h_{\times}(t) &= 0,
     \label{eq:hcross} 
\end{align}
where $A_{\text{fit}}$ is the amplitude, $t_0$ is the epoch at which the burst peak arrives the Earth, and $W$ is the duration of the burst. Strongly beamed in the direction of the cusp velocity is the GW burst, with cosmic string loops thought to generate cusps with high efficiency ($\mathcal{O}(1)$ per oscillation period). The orientation is assumed to be random and depend on the string configuration \cite{blanco1999form}. It is important to mention that $A_{\text{fit}}$ carries dimensions of sec$^{-1/3}$.The peak value at $t = t_0$, which is dimensionless, is defined as follows:
\begin{equation}
    A_{\text{peak}} \equiv h_+(t_0)=\left(\frac{1}{2}W\right)^{1/3}A_{\text{fit}}.
    \label{eq:Apeak}
\end{equation}

By substituting Equation \eqref{eq:hplus} into Equation \eqref{eq:rt}, the analytic expressions for the pre-fit timing residual caused by the GWCS are derived, using $A_{\text{fit}}$,
\begin{align}
    r(t) &= F^+(\hat{\Omega},\hat{p}) \times
    \begin{cases}
        0 &  (t<t_0-\frac{W}{2}) \\
        \begin{aligned}
            & A_{\text{fit}} \bigg[ \frac{3}{4} \bigg( \bigg( \frac{W}{2} \bigg)^{4/3} \mp \lvert t-t_0 \rvert^{4/3} \bigg) \\
            & \quad - \bigg( \frac{W}{2} \bigg)^{1/3} \bigg( t-\bigg( t_0-\frac{W}{2} \bigg) \bigg) \bigg] 
        \end{aligned} &  \left( t_0-\frac{1}{2}W \leq t < t_0+\frac{1}{2}W \right) \\
        -\frac{1}{4}(\frac{1}{2})^{1/3}A_{\text{fit}} W^{4/3}& (t\ge t_0+\frac{1}{2}W)
    \end{cases}.
    \label{eq:r(t)}
\end{align}
In the second line, the use of $\mp$ signifies that the $-$ sign is to be placed before $t_0$ and the $+$ after.
The effect of GWCS has been incorporated into the {\sc TEMPO2} timing model. This allows us to simulate residuals (or TOAs) of the GWCS. The updated timing model parameters are $(A_{\text{fit}}, t_0, W, \alpha, \delta, \zeta)$. Here, $\zeta$ is the principal polarization angle. This parametrization proves advantageous for the simulation of timing residuals induced by GWCS. Introducing a second parametrization of the GW burst proves advantageous for the search procedure. The GW burst in this parametrization is represented by two orthogonal components, $A_1$ and $A_2$ , with $A_1=A_{\text{fit}}$cos(2$\zeta$) and $A_2=A_{\text{fit}}$sin(2$\zeta$). These components are associated with the two polarization modes of the GW, and this parametrization allows for the search of all possible GW polarizations. {\sc TEMPO2} software package allows the use of the following parameters in its parameter files: GWCS$\_$A1, GWCS$\_$A2, GWCS$\_$POSITION, GWCS$\_$EPOCH, and GWCS$\_$WIDTH, which correspond to the cosmic string parameters $A_1$, $A_2$, sky position, burst epoch, and width, respectively.

\begin{figure}[htbp]
    \centering
    \hspace{-0.062\textwidth}  
        \includegraphics[width=0.765\textwidth]{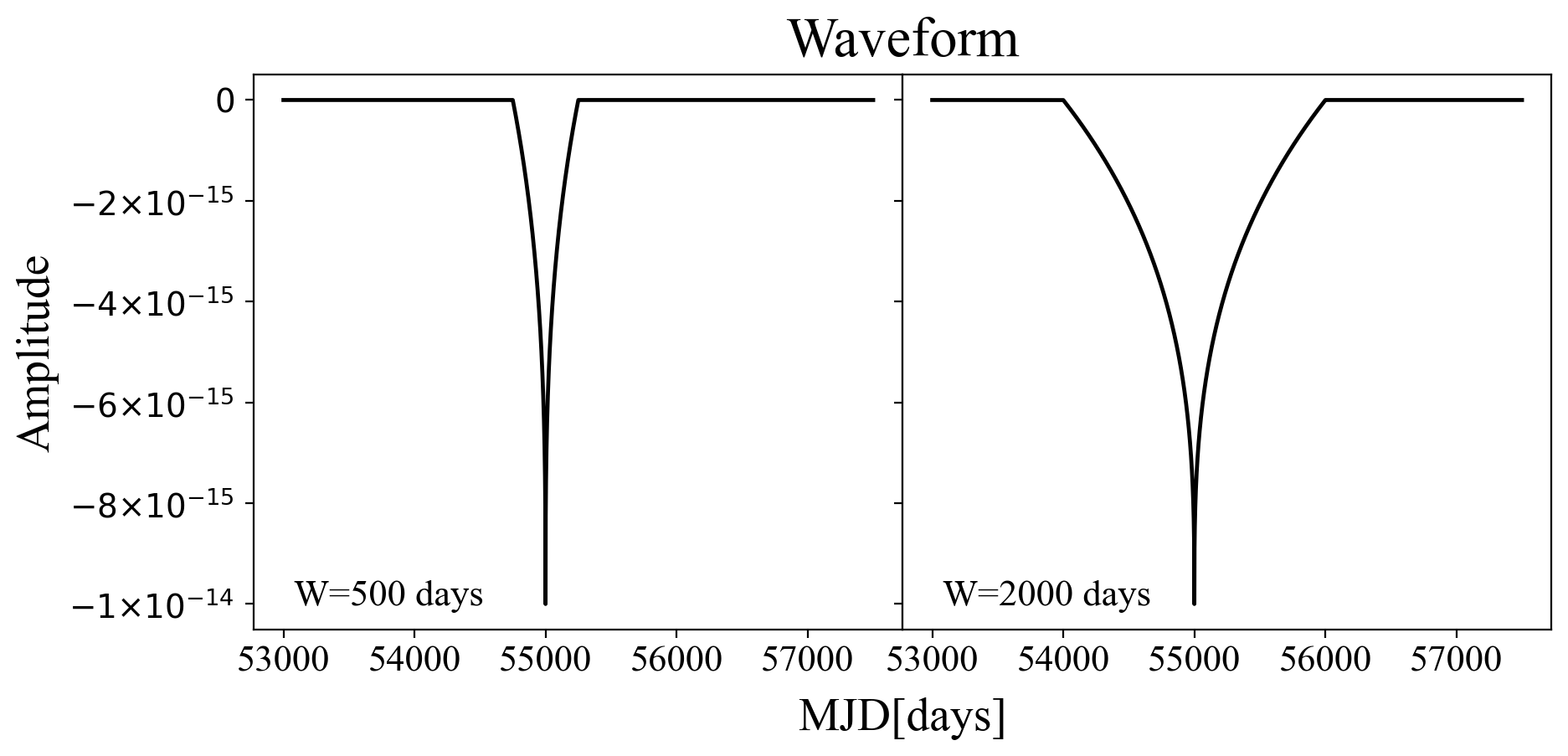}
    \vspace{0.5em}  

        \includegraphics[width=0.7\textwidth]{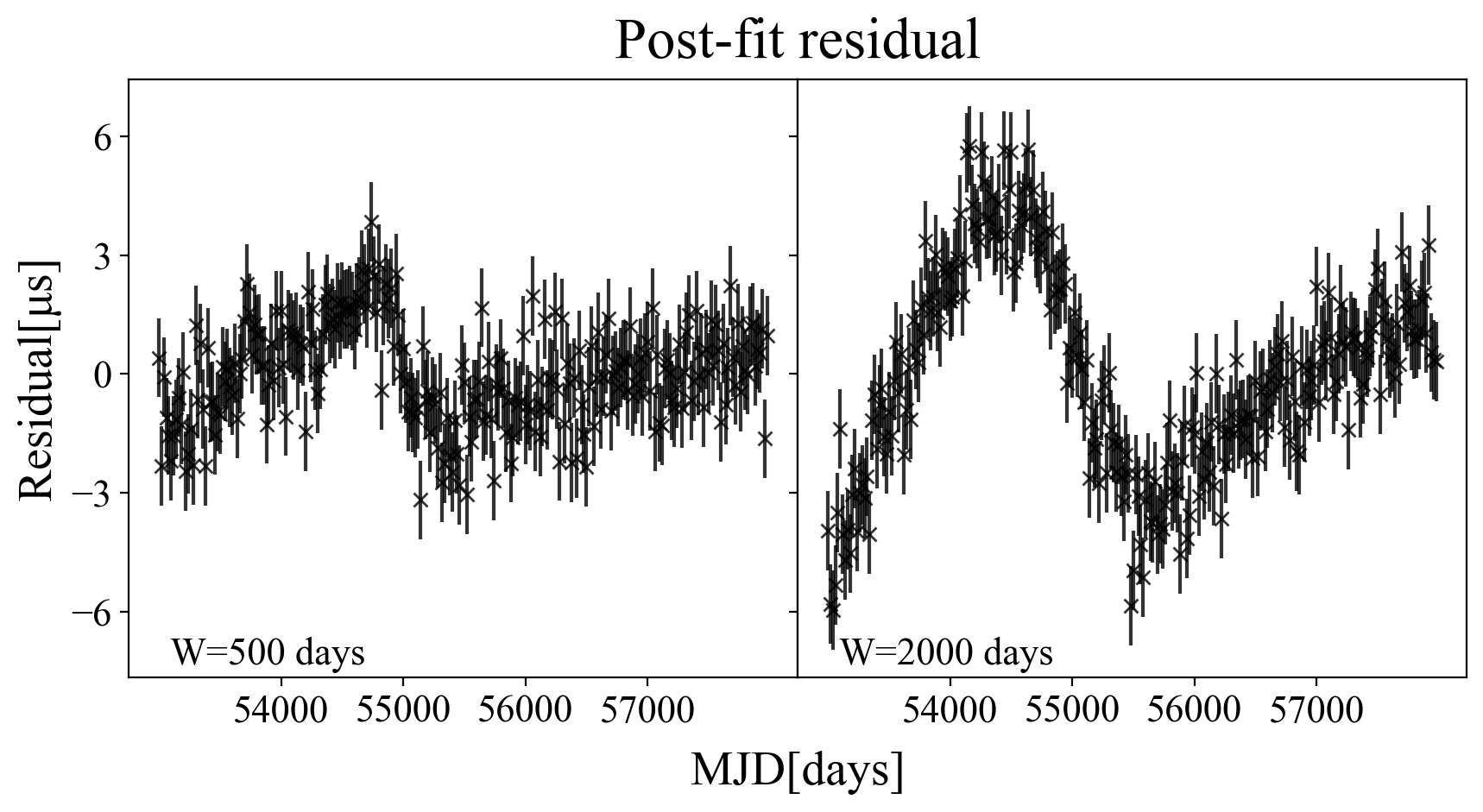}
    \caption{The top panels display the waveforms of GWCS. The simulated post-fit timing residuals, accompanied by 1 $\mu$s of Gaussian white noise, are displayed in the bottom panels. The gravitational wave amplitude is $A_{\text{peak}}=10^{-14}$, and the peak occurs at MJD 55000. These post-fit residuals are utilized to fit pulsar parameters like pulse period and spin-down rate, as shown in the bottom panel.}
    \label{fig:residual}
\end{figure}

Examples of the waveform and simulated timing residuals are shown in Figure \ref{fig:residual}. The {\sc PTASIMULATE} software package were used to simulate cosmic string burst events as described above. We assume that the position and polarization angle of the gravitational wave source are $(\alpha=57.3^{\circ}, \delta=-57.3^{\circ})$ and $\zeta=30^{\circ}$. Timing residuals are produced by infusing GWCS signal with an amplitude of $A_{\text{peak}}=10^{-14}$ at the center of the observation (MJD 55000) and 1 $\mu$s of Gaussian white noise using Equation (\ref{eq:r(t)}). The burst widths are set to 500 days (panel on the left side) and 2000 days (panel on the right side), corresponding to $A_{\text{fit}}=1.59\times 10^{-15}$ and $1.0\times 10^{-15}$, respectively.

Because of the crucial role of detecting a common spatially uncorrelated red noise process (CURN) identified by NANOGrav, PPTA, EPTA, and CPTA. Thus, a CURN with a specified spectral index is included in our model.
A power law can be used to model the CURN, with its power spectrum governed by two hyperparameters, $(A, \gamma)$ \cite{phinney2001practical}:
\begin{equation}
    P(f) = A^2 \left(\frac{f}{yr^{-1}} \right)^{-\gamma},
    \label{eq:P(f)}
\end{equation}
Here, $f$ denotes the frequency of the spectral component, $A$ represents the characteristic amplitude of the red-noise process  at a frequency of $yr^{-1}$ , and $\gamma$ is the spectral index associated with the characteristic amplitude. The spectral index of a stochastic gravitational wave background, arising from an ensemble of supermassive black hole binaries, is expected to be 4.33. However, the spectral index of the CURN was found to have a maximum a posteriori value of 5.5 \cite{arzoumanian2020nanograv}. In this paper, we present two sets of results for the CURN that utilize both of these fixed spectral indices ($\gamma_{\text{CURN}}=\text{4.33 and 5.5}$).

\section{Methods}
\label{sect:methods}
We use the Markov Chain Monte Carlo (MCMC) approach for the search of GWCS. In this section, we will provide a concise summary of the techniques. The methods employed in this search are detailed in Sun et al. \cite{sun2023implementation} and Aggarwal et al. \cite{aggarwal2020nanograv} to search for gravitational wave memory and continuous gravitational wave signals. The TOA residuals for an individual pulsar can be represented as a combination of multiple stochastic deterministic and stochastic processes:
\begin{equation}
    {\delta t} = {\delta t}_{\text{cs}}+{M\epsilon}+{Fa}+{F}_{\text{gw}} {a}_{\text{gw}}+{n}
    \label{delta_t}.
\end{equation}
Where ${\delta t}$ represent the residual time series for the pulsar. The term ${\delta t}_{\text{cs}}$ denotes the residuals induced by GWCS; ${M}$ is the design matrix representing the linearized timing model, which is responsible for accounting for the uncertainty in the residuals resulting from an imperfect timing model fit ${\epsilon}$; The design matrix ${F}$ corresponds to the pulsar's intrinsic Gaussian red-noise process, modeled as a Fourier series with coefficients represented by ${a}$; likewise, ${F}_{\text{gw}}$ and ${a}_{\text{gw}}$ are the design matrix and Fourier coefficients for the CURN, resepectively; lastly, the elements of vector ${n}$ represent uncertainty in the observed TOAs, which follow a Gaussian white-noise distribution.

Based on the estimations of the Gaussian process parameters, GWCS signal, and timing model parameters, it is possible to generate residuals ${n}$:
\begin{equation}
  {n}={\delta t} - {\delta t}_{\text{cs}}-{M\epsilon}-{Fa}-{F}_{\text{gw}} {a}_{\text{gw}}.
    \label{n}
\end{equation}

As the terms on the right-hand side are approximations, this constitutes merely an estimation of white noise. Nevertheless, assuming the white noise follows a Gaussian distribution, the likelihood of observing this specific sequence of white-noise residuals can be expressed as:
\begin{equation}
    p({n})=\frac{exp(-\frac{1}{2} {n}^T N^{-1} {n})}{\sqrt{2\pi detN}}.
    \label{p(n)}
\end{equation}
Above, $N$ is a covariance matrix that represents the white-noise uncertainties in each observed TOA, and ${n}^{T}$ is the transpose of ${n}$.

After the removal of deterministic effects, the remaining pulsar timing residuals have the same likelihood of following a Gaussian white noise distribution as the detection of a GWCS signal. This means that,
\begin{equation}
    p({\delta t} \mid {\epsilon}, {a}, {a}_{\text{gw}}, A_{\text{fit}}, t_0, W, \hat{{\Omega}},\hat{{p}},\zeta) = \frac{exp(-\frac{1}{2} {n}^T N^{-1} {n})}{\sqrt{2\pi detN}}.
    \label{p(t)}
\end{equation}

The parameter space dimensionality can be reduced through the analytical marginalization of the likelihood in Equation \eqref{p(t)} with respect to the parameters that describe the Gaussian processes \cite{lentati2013hyper, van2014new, van2015low}. The final marginalized likelihood obtained is
\begin{equation}
    p({\delta t}\mid A_{\text{fit}},W,t_0,\hat{{\Omega}},\hat{{p}},\zeta)=\frac{exp(-\frac{1}{2} {q}^T 
 {C}^{-1} {q})}{\sqrt{2\pi det{C}}},
\end{equation}
where the following definitions are provided:
\begin{equation}
    {q}={\delta t} -{\delta t_{\text{cs}}},
\end{equation}
\begin{equation}
    {C}=N+{TDT}^T,
\end{equation}
\begin{equation}
     {T}= \begin{bmatrix}
    {M} \hspace{3mm} {F}\\
    \end{bmatrix},
\end{equation}
\begin{equation}
    {D}=\begin{bmatrix}
    \infty \hspace{3mm} 0\\
    0 \hspace{3mm} \phi\\
    \end{bmatrix},
\end{equation}
where $\infty$ is a diagonal matrix of infinities, which effectively provides unconstrained priors on the timing model parameters. $\phi$ is a covariance matrix that represents the individual red noise and the CURN Fourier coefficients. As a result of employing a CURN, the $\phi$ matrices are diagonal, with each diagonal element representing the red-noise power at the corresponding frequency bin, as specified in Equation \eqref{eq:P(f)}. ${C}^{-1}$ is efficiently computed using the Woodbury matrix identity \cite{woodbury1950inverting}. In this identity, the matrix ${D}$ is present solely as an inverse. As a result, the diagonal matrix containing infinities is effectively replaced by a matrix of zeros during the likelihood calculation.

\begin{table}[H]
\centering
\begin{minipage}[]{150mm}
\caption{Prior distributions used in all analyses of this article, including red noise, CURN, GWCS.}
\label{tab:parameters}
\end{minipage}
\setlength{\tabcolsep}{1pt}
\small
\begin{tabular}{c@{\hspace{0.8cm}}c@{\hspace{0.8cm}}c}
  \hline\noalign{\smallskip}
Parameter &  Prior & Description \\
  \hline\noalign{\smallskip}
log$_{10}$A$_{\text{rn}}$ & LinearExp($-$17, $-$11) & Amplitude of intrinsic pulsar red noise\\
$\gamma_{\text{rn}}$ & Uniform(0, 7) & Spectral index of intrinsic pulsar red noise \\
log$_{10}$A$_{\text{CURN}}$ & LinearExp($-$17, $-$11) & Amplitude of GWB\\
$\gamma_{\text{CURN}}$ & Uniform(0, 7) & Spectral index of GWB\\
log$_{10}$A$_{\text{GWCS}}$ & LinearExp($-$18, $-$11) & Amplitude of GWCS\\
W$_{\text{GWCS}}$ & Uniform(0, 5000) & Width of GWCS\\
t$_{\text{GWCS}}$ & Uniform(MJD 53500, MJD 59100) & Epoch of GWCS\\
$\zeta_{\text{GWCS}}$& Uniform(0, $\pi$) & Polarization of GWCS\\
$\alpha_{\text{GWCS}}$& Uniform(0, $\pi$) & Polar angle of GWCS source\\
$\delta_{\text{GWCS}}$& Uniform(0, 2$\pi$) & Azimuthal angle of GWCS source\\
  \hline
\multicolumn{3}{p{0.99\linewidth}}{\footnotesize Notes: There are a total of six global GWCS parameters. Imposing priors on the logarithm of the amplitude is mathematically equivalent to applying uniform priors directly on the amplitude itself. Red noise and CURN prior parameters satisfy the power law distribution of Equation \eqref{eq:P(f)}.}
\end{tabular}
\end{table}

We favored ignorance priors for our model parameters and implemented uniform or log-uniform priors for all. We employed the same priors for noise parameters as Reardon et al. \cite{reardon2023gravitational} and Reardon et al. \cite{reardon2023search}. The Table $\ref{tab:parameters}$ provides the prior distributions of  model parameters in the Bayesian Search for Global GWCS using the full PTA that we are interested in. The Bayes factor, denoted as $\mathcal{B}_{\text{gw}}$, was employed as the detection statistic to compare the GW model with a noise-only model. Bayes factor was computed using the Savage-Dickey approximation, as described by Dickey \cite{dickey1971weighted},
\begin{equation}
    \mathcal{B}_{\text{gw}}=\frac{{\varepsilon}_{\text{gw}}}{{\varepsilon}_{\text{noise}}} \approx \lim_{A_{\text{fit}} \to 0} \frac{p(A_{\text{fit}})}{p(A_{\text{fit}}|\delta{t})}.
    \label{B}
\end{equation}
The evidence ratio (${\varepsilon}_{\text{gw}}/{\varepsilon}_{\text{noise}}$) for the GW and noise only models can be approximated as the ratio of the prior to posterior probability as the GW amplitude approaches zero. This computation is significantly more computationally efficient than a complete evidence integral, as it employs posterior samples that are located near the low amplitude prior boundary.

This likelihood calculation and these signal models are implemented in the {\sc ENTERPRISE} \cite{ellis2019enterprise} and {\sc ENTERPRISE$\_$EXTENSIONS} \cite{taylor2021enterprise_extensions}. The {\sc PTMCMCSAMPLER} \cite{ellis2017jellis18} package implements the MCMC sampler to extract samples from the posterior distributions.

\section{Results}
\label{sect:result}
\subsection{Earth-term GWCS Search}
\begin{figure}[htbp]
\centering
\includegraphics[width=0.9\textwidth]{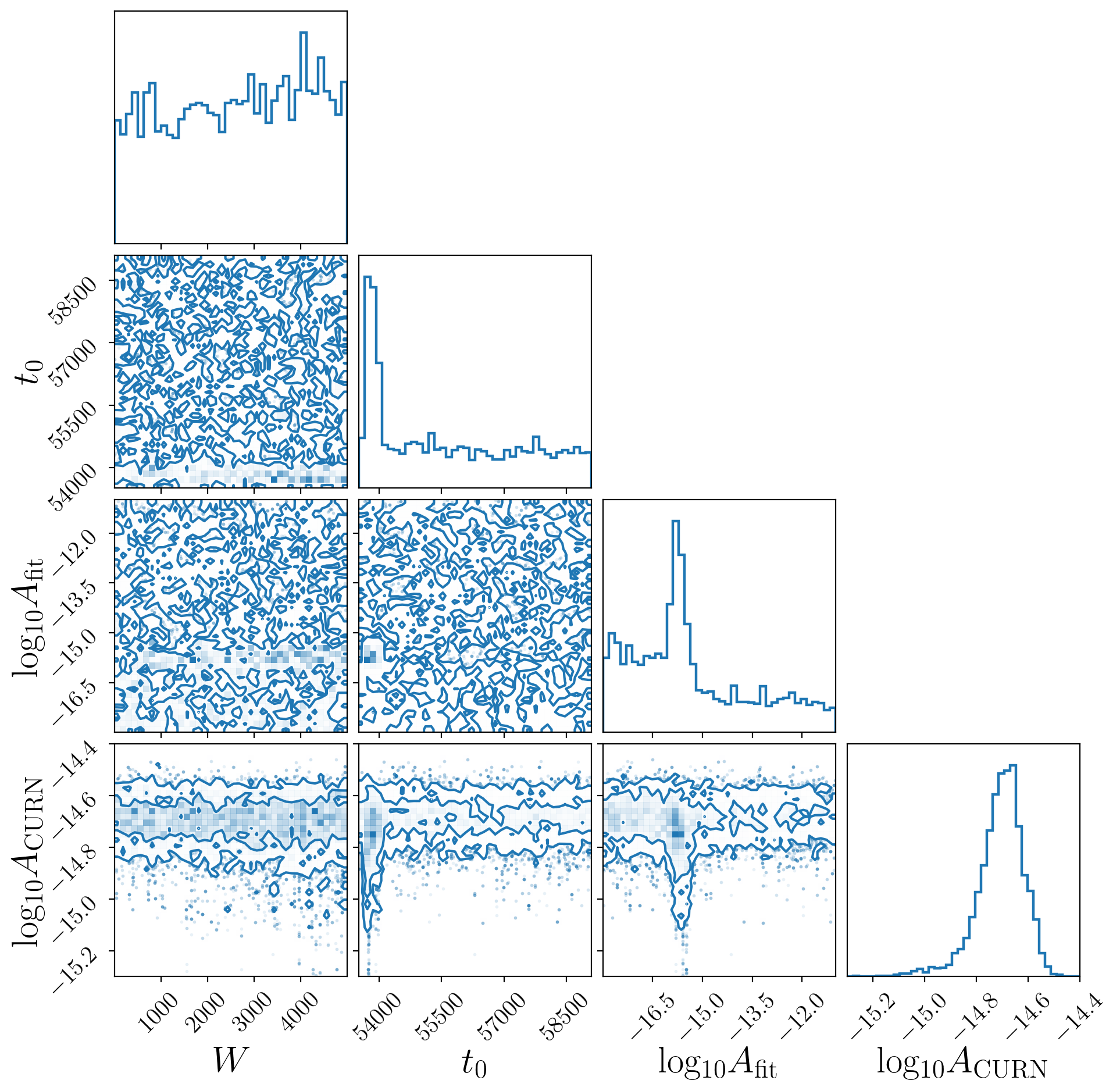}
\caption{A corner plot displaying the marginalized 1D and 2D posterior distributions for four essential model parameters: burst strain amplitude log$_{10}A_{\text{fit}}$,  burst epoch $t_0$, width $W$, and CURN amplitude log$_{10}A_{\text{CURN}}$.  In the presence of the GWCS model, the favorable localization of  log$_{10}A_{\text{CURN}}$ suggests that the CURN remains detectable. Moreover, the tail of log$_{10}A_{\text{fit}}$ stretches to a very small amplitude, suggesting that the model still strongly supports $A_{\text{fit}}\sim 0$.}
\label{earth-term}
\end{figure}

By applying MCMC sampling, we initiated a Bayesian search for GWCS in the Earth-term, contrasting two models: (1) a noise-only model and (2) a model that includes both noise and the GWCS signal. The noise-only model included white noise, achromatic red noise (Red), dispersion measure (DM), high fluctuation frequency (HFF), Chromatic (Chr), low-frequency band noise (BN), and a CURN process \cite{reardon2023search}. The signal model incorporated identical noise processes together with GWCS signal. We employed the product-space sampling technique \cite{10.1111/j.2517-6161.1995.tb02042.x} to sample both models simultaneously. This allows us to determine the posterior probability for the GWCS signal and computes the Bayes factor of $\mathcal{B}_{\text{gw}}=\text{0.7}$ for the GWCS signal model compared to noise only, this Bayes factor is too small to consider it detected. Figure \ref{earth-term} illustrates the posterior probability distributions for the GWCS signal as well as the global spatially uncorrelated red-noise process. 

By analyzing the posterior distribution, we can pinpoint a geographically significant areas of high activity near MJDs 54000. The feature near MJD 54000 is situated approximately in the beginning of our observations, during a period when there were significant gaps in data for multiple pulsars. At the beginning of our data set, there was a small number of pulsars being observed, and the observations exhibited less regularity. Data sparsity complicates the task of constraining any signal in these time. A heightened degeneracy is observed for events occurring earlier in the dataset when a quadratic pulsar timing model is applied to the pulsar's rotational frequency and its derivative. This implies that the signal model can exhibit consistency with a GWCS event of significant magnitude, which is appropriately eliminated by the marginalization of the timing model.

\subsection{Pulsar-term Upper Limits}
\begin{figure}[htbp]
    \centering
    \includegraphics[width=0.6\textwidth]{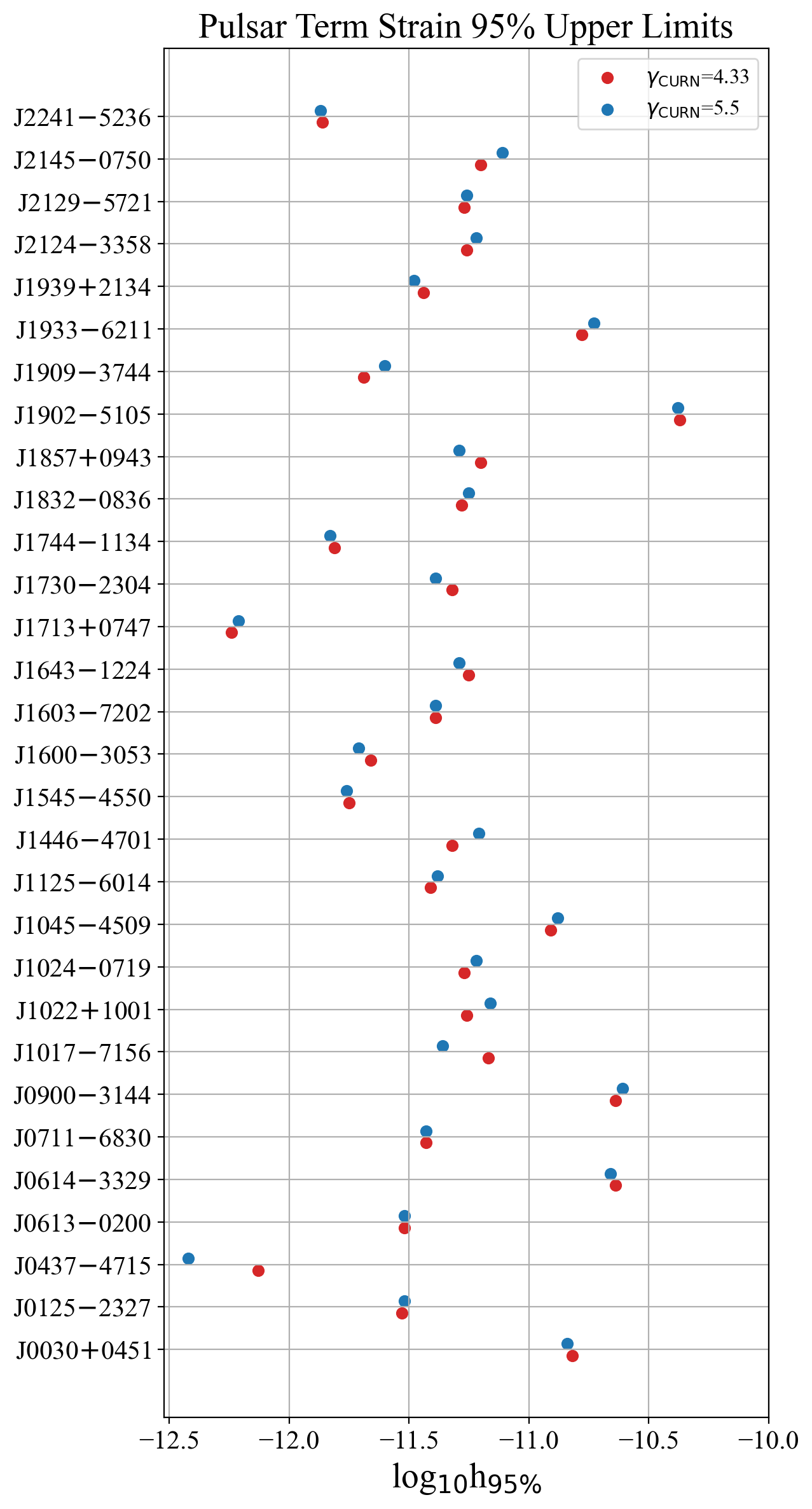}
    \caption{A plot of the upper limits of GWCS strain amplitude in the pulsar-term. In order to obtain the upper limits, we fixed duration of the burst $W$ = 100 days and obtained the upper limits for the pulsar-terms with fixed spectral indices $\gamma_{\text{CURN}}$ = 4.33 and 5.5, respectively. We observe minimal variation in the pulsar-term upper limits when different CURN spectral indices are fixed.}
    \label{pulsar-term}
\end{figure}

It is almost impossible to make a confident detection with a single pulsar term search of GWCS, but we could still use non-detection in the pulsar term search to set upper limits.
We report upper limits on the amplitude of the GWCS strain. The pulsar-term upper limits on GWCS are illustrated in Figure \ref{pulsar-term}, which employs both fixed spectral indices ($\gamma_{\text{CURN}}=\text{4.33 and 5.5}$). 
As the pulsar-term upper limits are calculated individually for each pulsar, any information about the signal's sky location is effectively lost. This means that it becomes challenging to distinguish between a weak GWCS event and one that originates from a sky location where the antenna pattern is weak relative to the position of the pulsar.

The upper limit $A_{\text{fit}}$ is determined by setting the width at $W=\text{100}$ days, and then we calculate the upper limit $h$ for the pulsar term using Equation (\ref{eq:Apeak}). It is evident that the pulsar-term upper limits are not significantly influenced by the selection of spectral index for the majority of pulsars.

\begin{figure}[H]
    \centering
    \includegraphics[width=0.9\textwidth]{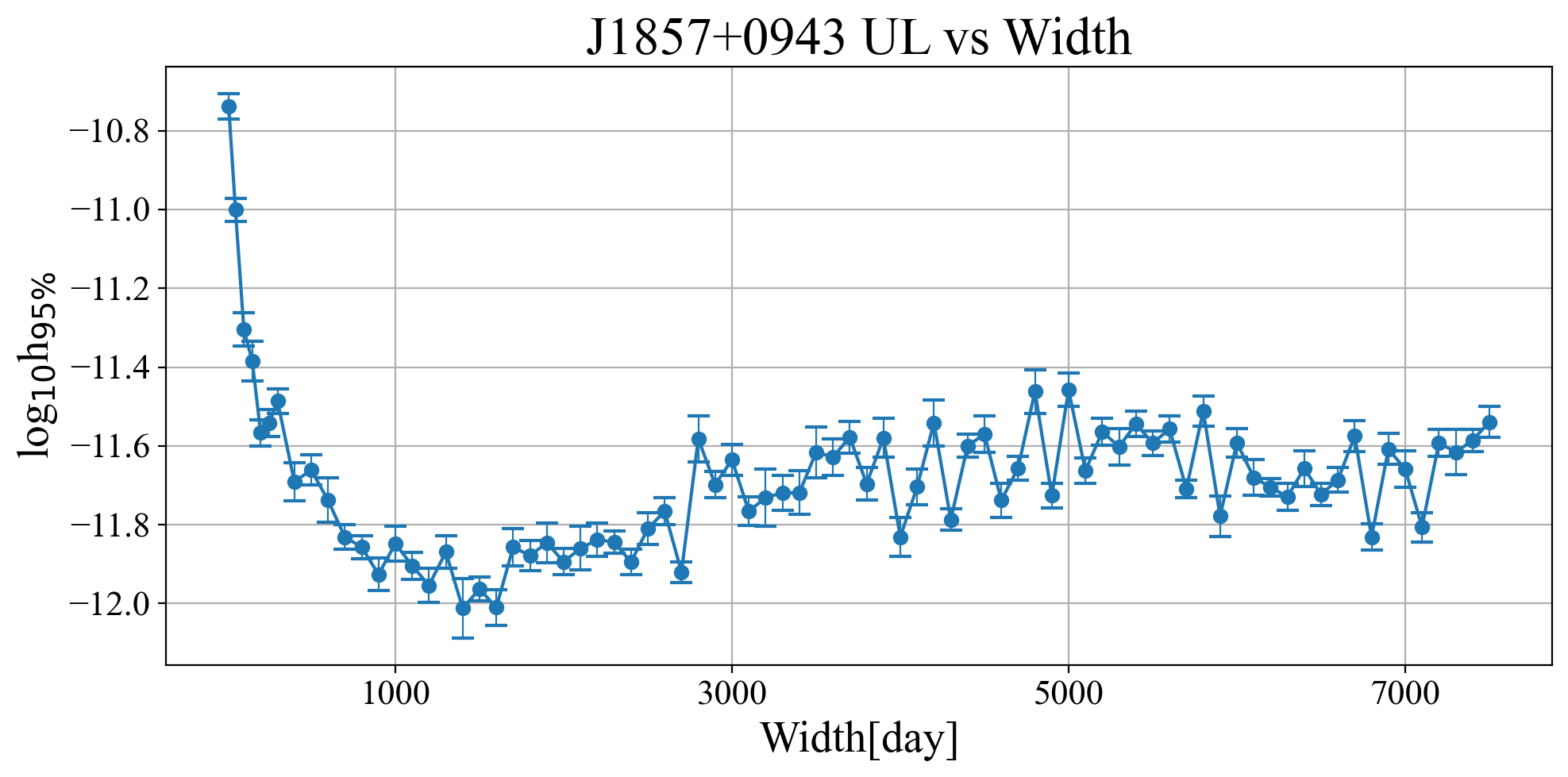}
    \caption{The relationship between the GWCS amplitude and burst width is constrained, assuming a fixed CURN spectral index of $\gamma_{\text{CURN}}=\text{4.33}$.}
    \label{h-width}
\end{figure}

The constraints on $h$ can be expressed as a mathematical function that depends on the epoch, width, and sky position. Only the width has a physical significance, which is equivalent to the loop size of the cosmic string. 
In Figure \ref{h-width}, we obtained the upper limit $h$ as a function of the width for  PSR J1857$+$0943. 
The early- and late-time bursts lack credibility due to insufficient data available prior to or following the occurrence of these bursts, which hinders the ability to accurately determine their amplitude. As the width increase, the constraint on $h$ becomes more stringent. Since the timing residuals caused by the GWCS increase as the event duration get longer. 
In Figure \ref{h-width}, the upper limit $h$ is a function of the width. As width increases, the strain amplitude $h$ decreases, then increases, and finally flattens out.The limitation is strongest when the width is about 1200 days and the upper limit is about $10^{-12}$.

\begin{figure}[h]
    \centering
    \includegraphics[width=0.9\textwidth]{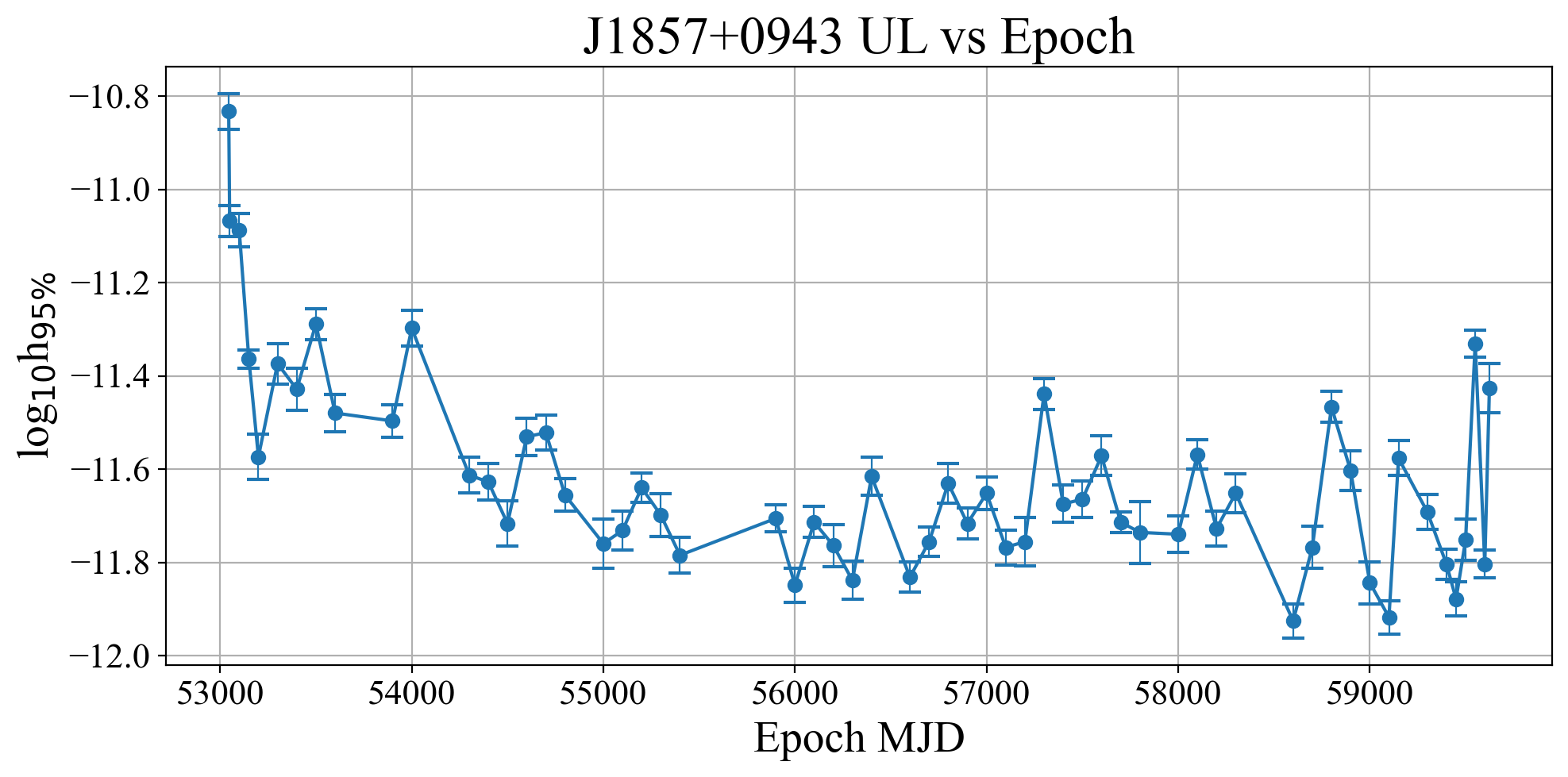}
    \caption{The upper limit on the GWCS strain amplitude as a function of burst epoch $t_0$ is derived by fixing the burst width at $W = 100$ days and adopting a CURN power-law spectral index of $\gamma_{\text{CURN}} = 4.33$.}
    \label{h-epoch}
\end{figure}

A term specific to each pulsar is introduced to impose a constraint on the upper time analysis for individual pulsars within the array. The upper bound h is a function of the burst epoch $t_0$ by fixing $W=\text{100}$ days for PSR 1857$+$0943 in Figure \ref{h-epoch}. When epoch $t_0$ is the median value, the upper limit of the amplitude stabilizes, and the constraint on h is the weakest at the beginning and end. When epoch $t_0$ is about 59000, the GWCS strain amplitude upper limit is the strongest because of a new ultra-wideband receiver, possibly.


{
Here we translate the upper bound on the peak amplitude shown in Figure \ref{h-width} into constraint on the cosmic string tension.
The upper bounds on the time-domain amplitude with different width shown in Figure \ref{h-width} can be translated into the Fourier strain amplitude $\tilde{h}_{\text{lim}}$. 
The expected number of gravitational wave bursts from cosmic string cusps reaching Earth today with amplitudes exceeding $\tilde{h}_{\text{lim}}$ is denoted as $N_{\text{GWCS}}$. Given the absence of detected cosmic string GW bursts with amplitudes above $\tilde{h}_{\text{lim}}$, a random Poisson process analysis at the 95\% confidence level excludes scenarios where $N_{\text{GWCS}} > 2.996$. Using this constraint, we establish an upper limit on the cosmic string tension. The detail of the translation can be found in \cite{yonemaru2021searching}. }

\begin{figure}[h]
    \centering
    \includegraphics[width=0.7\textwidth]{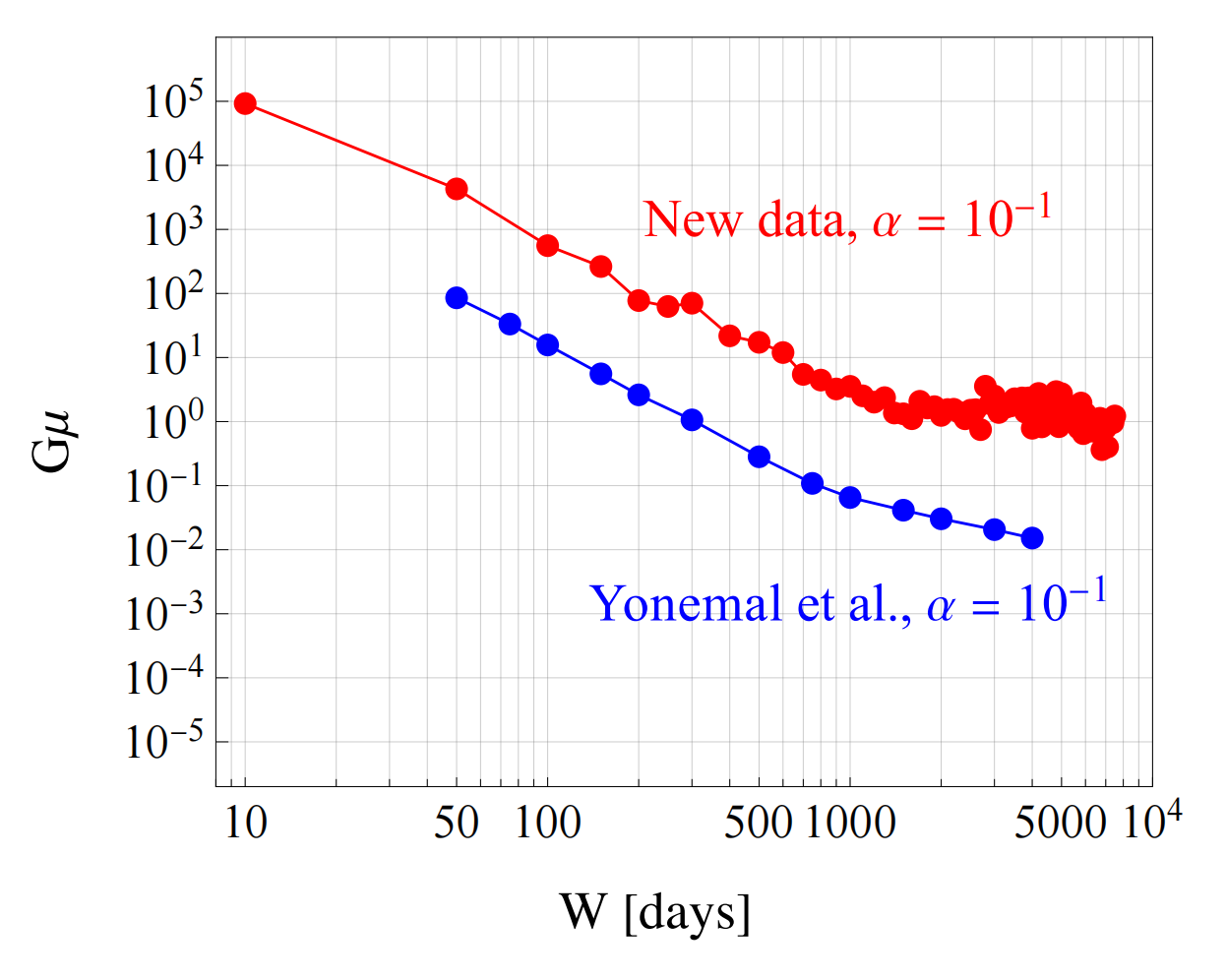}
    \caption{Constraint on $G\mu$ is derived for different widths, with the coupling constant $\alpha$ fixed at 0.1.}
    \label{Gmu-w}
\end{figure}

{Figure \ref{Gmu-w} presents the constraint on the cosmic string tension as a function of burst width 
W, derived from the single pulsar dataset of PSR 1857+0943, alongside the results from \cite{yonemaru2021searching}. The $\alpha$ in Figure \ref{Gmu-w} represents the size of the primordial string loop, and here we compare the cosmic string tension limit specifically for $\alpha$ = 0.1. As shown in Figure \ref{Gmu-w},the upper bound becomes more constraining for larger W due to two factors.
First, as shown in Figure \ref{h-width}, the upper bound on improves for larger W. Second, larger loops produce stronger GW amplitudes but have lower number densities. While the former enhances detectability, the latter reduces it. Considering both effects, we find that the former effect dominates and larger loops,  equivalentlylarger W, easier to detect and leading to tighter constraints.}
{However, the upper limit from \cite{yonemaru2021searching} is more than an order of magnitude tighter than our results. This discrepancy arises because we use only the dataset of a single pulsar (pulsar term), whereas Yonemaru et al. \cite{yonemaru2021searching} utilized the full PPTA dataset (Earth term).Additionally, MCMC sampling is very computationally intensive and time-consuming. In future work, we plan to extend our analysis to include the Earth term search.}

{So far, only Yonemaru et al. \cite{yonemaru2021searching} and this work have studied individual GW bursts from cosmic string cusps using PTA. In addition, cosmic string parameters can also be constrained by the cosmic microwave background (CMB) \cite{2014planck} and the stochastic GW background. While the stochastic background provides a stronger constraint on string parameters, we emphasize that GW burst searches offer independent constraints.
The future SKA telescope will observe a large number of millisecond pulsars with high timing precision. With the improved sensitivity of SKA, the prospects for detecting single GW bursts will be significantly enhanced. A stringent constraint on the cosmic string tension from SKA would be highly valuable for testing various models of cosmic superstrings, such as the KKLMMT model \cite{2003kachru}.}


\section{Conclusions}
\label{sect:conclusions}
In this paper, we present the first Bayesian searh for the GWCS using PPTA-DR3, there is no significant evidence of GWCS. Therefore, we have place upper limits on the strain amplitude of pulsar-term GWCS for the 30 millisecond pulsars. We find that incorporating a CURN with various spectral indices into the noise model has a negligible impact on the upper limits. And the upper limit range of the amplitude of the pulsar-term GWCS is concentrated between $10^{-12}$ and $10^{-11}$. 
By analyzing an individual pulsar PSR J1857$+$0943, we place upper limits on the amplitude of GWCS events as a function of width and event epoch.{We obtain the upper limit on the cosmic string tension as a function of burst width and will present the results of all-sky Earth-term upper limits in future work. With its improved sensitivity, SKA will place significantly tighter constraints on cosmic string parameters, making it a valuable tool for testing cosmic string models. Moreover, such constraints will provide an independent probe for investigating the clustering of cosmic string loops in our Galaxy, which cannot be tested through CMB analysis or stochastic GW background.}

\vspace{4em}

\noindent \textbf{authorcontributions:} software, Y.X. and Y.-R.W.; investigation, J.-B.W., S.K., Y.F., S.M., C.-J.R., S.-Q.W. and D.Z.; data curation, Y.X., Y.-R.W., D.-J.R. and A.Z.; writing---original draft preparation, Y.X.; writing---review and editing, J.-B.W., W.-M.Y. and A.K.; visualization, Y.X., S.K. and J.Z.; funding acquisition, J.-B.W., W.-M.Y., V.-D.M and X.-J.Z. All authors have read and agreed to the published version of the manuscript.

\vspace{1em} 

\noindent \textbf{funding:} This research was funded by the Major Science and Technology Program of Xinjiang Uygur Autonomous Region (No.2022A03013-4), the Zhejiang Provincial Natural Science Foundation of China (No.LY23A030001), the Natural Science Foundation of Xinjiiang Uygur Autonomous Region  (No.2022D01D85), National Natural Science Foundation of China (No.12041304); W.M.Y. is supported by the National Natural Science Foundation of China (NSFC) project (No.12273100, 12041303), the West Light Foundation of Chinese Academy of Sciences (No.WLFC 2021-XBQNXZ-027), the National Key Program for Science and Technology Research and Development and the National SKA Program of China (No.2022YFC2205201, 2020SKA0120200); V.D.M. is supported via the Australian Research Council (ARC) Centre of Excellence CE170100004 and CE230100016, and receives support from the Australian Government Research Training Program; X.J.Z. is supported by the National Natural Science Foundation of China (Grant No.12203004) and by the Fundamental Research Funds for the Central Universities.

\vspace{1em} 

\noindent \textbf{dataavailability:} Observational data used in this paper are quoted from the cited works. Additional data generated from computations can be made available upon reasonable request.

\vspace{1em} 

\noindent \textbf{acknowledgments:} The Murriyang Parkes 64 m radio telescope, part of the Australia Telescope National Facility (\url{https://ror.org/05qajvd42}), is funded by the Australian Government and operated as a National Facility under CSIRO’s management. We acknowledge the Wiradjuri People as the Traditional Custodians of the land upon which the Observatory is located. We also pay respect to the Wurundjeri People of the Kulin Nation and the Wallumedegal People of the Darug Nation, the Traditional Owners of the land where this research was primarily carried out. Our heartfelt thanks go to the Parkes Observatory staff for their steadfast support over nearly twenty years. We are also thankful for the access provided by the CSIRO Information Management and Technology High Performance Computing group to the Petrichor cluster and their technical assistance.
\vspace{1em} 

\noindent The following abbreviations are used in this manuscript:\\
\noindent 
\begin{tabular}{@{}ll}
PTA & Pulsar timing array\\
GWs & Gravitational waves\\
GWCS & Gravitational-wave bursts from cosmic string cusps\\
CURN & Common spatially uncorrelated red noise 
\end{tabular}


\end{document}